\documentstyle[12pt,aasms4]{article}
\begin{document}
\title{Absorption of the Linear Polarization of the Galactic Background 
Radiation by the $\lambda$21-cm Line of HI}
\author{John M. Dickey}
\affil{Australia Telescope National Facility, PO Box 76, Epping, NSW 2121,
Australia,
and Department of Astronomy, University of Minnesota, 
116 Church St. SE, Minneapolis, MN 55455} 
\authoraddr{Department of Astronomy, 116 Church St. SE, Minneapolis, MN, 55455}

\begin{abstract}
Absorption lines at $\lambda$ 21-cm are detected in the Stokes Q and U 
components of the Galactic synchrotron background.  The lower limit
distance implied for the emission region is 2 kpc in the direction
(l,b) = (329.5$^{\circ}$,+1.15$^{\circ}$).  The Australia Telescope Compact
Array has the capability of mapping this absorption over large areas
of the Galactic plane.  Observations like these have the potential to
reveal the three dimensional structure of the Galactic synchrotron
emission throughout the Milky Way disk.
\end{abstract}

\section{Introduction} 

Spiral galaxies show diffuse non-thermal emission at meter- and
centimeter-waves from synchrotron emission by high energy
electrons interacting with the interstellar magnetic field.
This emission is intrinsically linearly polarized, with the 
electric vector perpendicular to the direction of the magnetic
field projected on the sky. Mapping the polarization of this 
non-thermal radio emission from nearby spiral galaxies gives an
image of their large-scale magnetic field structure (Ehle et al.
\markcite{Ehle96}1996,
Klein et al. \markcite{Klein96}1996, 
Beck et al. \markcite{Beck96}1996 and references therein).
In the Milky Way we might hope for a much more detailed picture
of the magnetic field configuration to come from observations of
the polarization of the non-thermal Galactic background radiation.
At high galactic latitudes this method gives good results, which
agree qualitatively with other tracers of the interstellar magnetic
field such as the polarization of starlight (Spoelstra,
\markcite{Spoelstra84}1984 and
references therein; Salter and Brown,\markcite{Salter88}
1988 and references therein).  At low
latitudes the polarization of the galactic background presents
a confused picture which has not yet been interpreted overall.
The recent surveys by Junkes, F\"{u}rst and Reich (\markcite{Junkes87}1987)
and by Duncan et al. (\markcite{Duncan97}1997) 
show how complex the distribution of 
polarization appears at low latitudes, even when observed by 
the relatively large beam of a single dish telescope.

One of the difficulties of interpreting the polarization of the
non-thermal Galactic background comes from Faraday rotation.
Direct measurement of Faraday rotation toward pulsars and extragalactic
continuum sources provides one of the most effective tracers of the
Galactic magnetic field structure at low latitudes (Rand and Lyne
\markcite{Rand94}1994, 
Han et al. \markcite{Han97}1997, 
Beck et al. \markcite{Beck96}1996).  For extended structure, however,
small scale variations in the Faraday rotation can cause small scale
fluctuations in the position angle of the polarization which may
be blended by the relatively large beam of a single dish telescope.
Small scale structure in the intrinsic position angle of the polarized
emission can be blended in the same way, so that the beam 
average polarized intensity may thus be much less than 
the intensity present on smaller angular scales (``beam depolarization'').
An interferometer generally resolves away most of the unpolarized
(Stokes I) emission, since it is smoothly spread over angles of
a few arc minutes or larger, but small angle variations in the position angle
of the polarized intensity will cause strong ``fringes'' in Stokes
Q and U, making the fractional polarization seem anomalously high,
sometimes greater than 100\%.
An example of this effect was discovered by
Wieringa et al. (\markcite{Wieringa93}1993)
with the Westerbork Synthesis Radio Telescope.  The lesson from 
that work is that aperture synthesis telescopes
can detect small scale features in the Stokes Q and U components of the
Galactic background which may be completely undetectable in Stokes I.
Unlike the large scale features mapped by single dish telescopes at
high latitudes, at low latitudes such small angular size 
structures in the linear polarization could originate in quite 
{\bf distant} emission regions.

A powerful method for determining distances to Galactic plane objects
is by measuring the velocities of absorption lines at $\lambda$21-cm.
Distances to the intervening HI clouds that cause the absorption can
be estimated from their velocities using a rotation model of the Milky
Way (see Burton, \markcite{Burton88}1988).  
In order to understand the origin of the small
scale structures in the Galactic linear polarization it would be useful to
measure spectra of the 21-cm line in absorption toward this emission.

All past studies of 21-cm absorption have used Stokes I emission for
the background continuum.  Generally, compact continuum sources must be used
because the strong, diffuse line emission masks the absorption, requiring 
subtraction of the ``off source'' spectrum, either explicitly or by the spatial
filtering effect of an aperture synthesis telescope (e.g. Dickey and Lockman,
\markcite{Dickey90}1990, and references therein).  
The 21-cm line does not show Stokes Q and
U emission, so confusion with line emission is not a problem for 
absorption of linearly polarized continuum emission, unless the 
telescope does not adequately reject the Stokes I spectrum from the
measured Stokes Q and U spectra due to instrumental polarization.
The required rejection is roughly a factor of
one in 1000, which is difficult to achieve, particularly for single 
dish telescopes.

In this project I have used the Compact Array of the Australia Telescope
National Facility to observe the Stokes Q and U components of the Galactic
non-thermal emission in a region at Galactic longitude 329.5$^{\circ}$,
latitude +1.2$^{\circ}$. 
%The field was selected from a mosaic survey of
%the area around {\it l}=330$^{\circ}$, b=+1.5$^{\circ}$.  
The 21-cm line is 
seen clearly in absorption, which allows us to place a lower limit of 
about 2 kpc for the distance to at least some of the polarized emission. 
The observations are described in section 2,
and the results discussed in section 3.  

\section{Observations}

The Australia Telescope Compact Array (ATCA) has capabilities for measuring
polarization which are superior to most other telescopes. 
The antenna and feed design is symmetric in reflection in azimuth
to reduce instrumental polarization. Its feeds 
are linearly polarized, which facilitates the calibration of the leakage
terms between the polarizations and thus improves the quality of the 
Stokes Q and U measurements.  Typical values for the leakage (calibrated
using 1934-638) are less than 3\%; more important, the leakages are 
constant to better than 0.1\% over several weeks.  This leads
to a rejection factor of less than 0.1\% on-axis, i.e. the fraction of
Stokes I flux from a point source at the field center which spuriously
appears as a Stokes Q, U, or V signal is less than $10^{-3}$.  This factor
degrades off-axis, so that a source 30' from the field center may
show as much as 0.5\% spurious linear polarization at $\lambda$20-cm.
For this reason mosaic mapping of Stokes Q and U should be done
with smaller spacing between pointing centers than is necessary
for Stokes I alone.
All calibration and reduction of these data used the MIRIAD
package.  Analysis and display of the results used the AIPS package.

This project was done in two steps.  The first was a mosaic map of the region
of strong linear polarization found by Duncan et al. 
(\markcite{Duncan97}1997) in the area
$326^{o} < l < 333^{o}, 0^{o} < b < 3^{o}$ (190 pointings on a square grid
of 20' steps in l and b).  This mosaic was made using the
210m configuration of the array in December, 1996.  The correlator configuration
(``Full\_128\_2'') gives two bands of 128 MHz each, with full Stokes 
parameters.  Each 128 MHz band is broken into 32 channels of 4 MHz.  The
two bands were centered at 1380 MHz and 1670 MHz.  Many channels were
rejected due to interference, but there remain more then 50 MHz in each
band which was quite clean.  Figure 1 shows both the Stokes I emission
in the mosaic area, and the polarized intensity over the area.
Some of the structures in the Stokes Q and U emission do not correspond 
with any discernable Stokes I sources.  These are similar to the 
features of Wieringa et al. 
(\markcite{Wieringa93}1993); they are almost certainly caused by
small scale variations in the position angle of the linear polarization, 
which show up at the relatively high spatial frequencies sampled by
the interferometer, whereas the Stokes I emission is more smoothly 
distributed and so resolved away by the ATCA.

For the second step, I selected a region of strong linear polarization and
observed a single beam area for 16 hours using a high resolution
correlator configuration (``Full\_4\_1024\_pol'').
This gives full Stokes parameters over a 4 MHz band which
is broken into 1024 channels separated by 3.91 kHz 
(0.82 km s$^{-1}$ at 1420 MHz, rebinned to 2 km s$^{-1}$ during processing).
The pointing center chosen for this second step was 
($\alpha,\delta$ [J2000]) = (15:56:51.10, -51:53:37.8) or 
(l,b) = (329.47,+1.16).  Further mosaic observations and 
more extensive processing after the observations show that there
are stronger linear polarization peaks in the region, so this field is not 
a particularly special one.  The Stokes Q and U maxima are roughly
80 mJy per beam in this region, and the linearly polarized intensity, PI = 
$\sqrt{ Q^{2} + U^{2} - \sigma_{noise}^{2}}$ is about 100 mJy per beam,
where $\sigma_{noise}$ is the rms noise in the map.  After tapering, 
the beam size was about 7.5'x5.2', so the conversion
factor between Jy per beam and Kelvins of brightness temperature is 
4.4 K/(Jy per beam).  Thus the peak polarized brightness temperature is about
0.4 K.  This is considerably stonger than the 100 to 150 mK polarized
intensity seen by the Parkes beam at 2.4 GHz in this area, though much
less than the 8 K seen by Wieringa et al. at 327 MHz in their strongest
area.

A grey scale map of the polarized intensity 
is shown on figure 2.  This was made using a line-free 1 MHz 
portion of the 4 MHz narrow band data (1418.9 - 1419.9 MHz).  Spectra
of the polarized intensity, $PI$ are shown
for each of 16 areas of 5' square (about one synthesized beam area each)
in the central region of the primary beam.
The polarized intensity averaged over each square
is reflected in the baseline intensity of each spectrum.  The best
absorption is obtained in the north-central part of the map,
where the continuum polarized intensity reaches over 100 mJy per beam.  
In this region the deepest absorption is in the velocity
range -25 to -40 km s$^{-1}$.

Statistics of $PI$ can be deceptive, since this is
a positive definite quantity in which a varying rms noise
across the band can mimic an emission feature.  In fact the
noise level does vary across the band, since the Stokes I HI emission 
detected in the primary beam is strong enough to nearly 
double the 40K system temperature. 
The rms noise spectrum in Stokes V gives a good estimate of
the noise as a function of frequency in Stokes Q and U as well.
This is 7 mJy rms off-line, rising to 10 mJy over the velocity
range -120 to +50 km s$^{-1}$.  This excess noise is coming entirely
from Stokes I, so it will enter the spectra on figure 2 without
any correlation with the continuum polarized intensity shown
in the grey scale.  It will be reflected in $PI$ both as 
increased fluctuations, and as an increased mean level in this
velocity range.  

In order to avoid the unpleasant statistical properties of the
$PI$ spectrum,
we can look directly at the spectra of Stokes Q and U.
The noise in these quantities remains zero mean, so that
spurious emission features should not be created by the
varying noise level across the band.  This property is preserved
even if we rotate the axes of Q and U by angle $\Theta$ 
to new axes, Q' and U', 
thus Q' = Q cos$\Theta$ - U sin$\Theta$ and
U' = Q sin$\Theta$ + U cos$\Theta$.
To best search for absorption of the polarized
continuum, we choose for $\Theta$ minus two times
the position angle of the polarization
measured 2 MHz off-line, i.e. $\Theta = -2\Psi = -\arctan\frac{U_{c}}{Q_{c}}$,
where $U_{c}$ and $Q_{c}$ are the Stokes parameters measured in the offline
continuum band.  (Faraday rotation over such
a small frequency offset is generally less than 10$^{\circ}$ 
or so.)  With this choice of $\Theta$ the new
Q' contains all the linearly polarized background continuum flux
but has no systematic bias due to noise.  Since the observed position angle
is changing across the map, due in part to the absence of the
zero spacings in the aperture synthesis and in part to real variations
in the position angle, this rotation must be done for each pixel of the
cube separately.

Spectra of Q' are shown in figure 3 for two areas with strong
polarized continuum on figure 2 (the second and third boxes in the second row).
The deepest absorption lines are centered at -36 and -28 km s$^{-1}$; 
in both spectra the absorption at this velocity is very deep, 
peak optical depths are in the range 1 to 1.6. The 
narrowest line has velocity width $\sigma_{v} \approx$ 2.5 km s$^{-1}$ or less.
There is a tentative absorption line near zero km s$^{-1}$, and
there is an apparent {\bf emission} line at -66 km s$^{-1}$.
If it is real, this must be in fact an absorption line which
intervenes between two regions of linear polarization with
quite different position angles (see next section).

\section{Discussion}

The strongest and most convincing absorption
lines on figure 3 are those centered
at -28 and -36 km s$^{-1}$.  Assuming that these
velocities are primarily due to Galactic
rotation, the implied kinematic distances are 2.0 and 2.5 kpc using 
the rotation curve of Brand and Blitz 
(\markcite{Brand93}1993), or 2.3 and 2.9 kpc using the simple
Camm's Law approximation
with Oort's Constant A $\equiv$ 14 km s$^{-1}$ kpc$^{-1}$ (Burton
\markcite{Burton88}1988).
Alternatively, if the cloud is beyond the sub-central point,
then the implied distances are 12.6 and 12.1 kpc.
Random velocities of HI clouds are typically 6 km s$^{-1}$, so the error 
on these distances is about 20\%.  The weaker line near zero km s$^{-1}$ is
probably due to local gas, if it is real.  

The most interesting question for
the polarized background is whether the positive line at -66 km s$^{-1}$
is real.  This is at about the 4-sigma level, so we
cannot give it high confidence, although it appears in several positions 
on the map (figure 2).  If it is real, it would imply distances of
either 4.0 or 10.1 kpc.  Even at 10 kpc
distance the line of sight is only 200 pc above the plane, so this is well
within the thick disk of synchrotron emission.

Absorption lines in Stokes Q and U can take on very unconventional
forms, including positive lines which mimic emission, very deep lines
that go well below the zero point of the background, and lines which
only change the position angle of the continuum, and not the intensity.
The polarized continuum is emitted in various regions along the line of sight.
The linear polarization of these various
contributions adds together more or less 
as a random walk on the Q-U plane, depending on the coherence of the magnetic
field on large scales.  A long line of sight across the inner galaxy must 
traverse regions of reversal of the large scale field, as well as many small
scale changes of the random field component.  In addition, intervening regions
of high electron density cause Faraday rotation of the emission from 
behind them; the change in $\Psi$ can be a large angle at 
$\lambda$21-cm.  For comparison, in the field shown on figure 1 there are
several pulsars with known rotation measures, varying from +9
to -560 radians m$^{-2}$.  Not having the zero-spacing information, 
it is not possible to map the rotation measure over the field, but
undoubtedly there are large variations on a range of angular scales.

An absorption line from an intervening HI cloud can affect the 
intensity and position angle of the net linear polarization in a variety
of ways.  In the case of significant internal Faraday
dispersion along the line of sight (Ehle \markcite{Ehle93}and Beck 1993,
Burn \markcite{Burn66}1966, Spoelstra \markcite{Spoelstra84}1984),
an absorption line near the front side of the emission region
has the effect of decreasing the Faraday depth,
and so increasing the polarization.  This could be the
situation for the line at -66 km s$^{-1}$, which appears positive.  
If there are two regions of polarized emission at different
distances along the line of sight whose position
angles $\Psi$ differ by nearly 90$^{\circ}$, then an 
absorption line from a cloud between the two can be either
positive or very negative, i.e. well below zero, depending on 
which of the two has stronger polarized intensity.
In the same way, it is possible to see spectral lines in Stokes Q and U
even in directions where the continuum polarized intensity is
zero, if the line of sight contains several emission regions
whose polarization vectors roughly cancel and there are optically thick
HI clouds between them.  This may be the
reason why the line at -66 km s$^{-1}$ appears with roughly
equal strength in some positions on figure 2 where the continuum level has
dropped to less than 50 mJy per beam.
Since the noise level in the spectra on figure 3 is high enough that
only the lines at -36 and -28 km s$^{-1}$ are detected above 5 $\sigma$,
and these lines
look like conventional absorption lines of high optical depth, I must defer
claims for such unconventional absorption lines as that at -66 km s$^{-1}$ 
until more observations can be done.

In addition to the possibilities for
determining distances to the various contributions
to the polarized continuum background, $\lambda$21-cm absorption offers the 
possibility of mapping the angular distribution of the cool HI gas on scales
of many arc minutes or even degrees.  Figure 2 shows that the narrow, deep
absorption line at -28 km s$^{-1}$ extends over several synthesized beam areas, 
but does not cover uniformly the entire 30' primary beam area.  Thus the 
cool cloud must be larger than $\sim$ 3 pc but smaller than $\sim$ 20 pc.  
The spatial structure of the absorption is reminiscent of 
maps of 21-cm absorption toward extended, bright continuum sources such as
HII regions and supernova remnants (e.g. Troland, Heiles and Goss,
\markcite{Troland89}1989; 
Liszt, Braun, and Greisen \markcite{Liszt93}1993).   In the future, this
type of observation may provide large scale maps of the atomic gas in
cool interstellar clouds, for comparison with CO maps and other tracers of the 
dense phases of the interstellar medium.

\acknowledgments

I am grateful to the staff of the ATNF for their hospitality and assistance,
and particularly to Bob Sault and Ron Ekers, without whose advice and 
encouragement this work would not have been possible.  
I am also grateful to Jim Caswell, Mark Wieringa, Mattias Ehle, 
Neil Killeen, and Roy Duncan for comments on the manuscript.
This research was supported in part by National Science Foundation
Grant 92-22130 to the University of Minnesota.  The ATCA is part of
the Australia Telescope which is funded by the Commonwealth
of Australia for operation as a National Facility managed by CSIRO.

\subsection*{Figure Captions}

Figure 1.  The Stokes I emission in the mosaic field (top) and the
polarized intensity, $PI$ (bottom).  Many structures 
corresponding to HII regions and supernova remnants are apparent in Stokes I, 
particularly near b = 0$^{\circ}$ (see Whiteoak and Green, 1996).  Some of
these are prominent also in linear polarization.  There are also regions of
relatively strong $PI$ which do not correspond to discrete sources in Stokes I.
The circle indicates
the primary beam area mapped with high frequency resolution to search for
21-cm absorption (figure 2).

Figure 2.  A grey-scale image of the polarized intensity, with
spectra averaged over a grid of 5' x 5' cells.  The grey
scale runs from 10 to 95 mJy beam$^{-1}$. The spectra are all on 
the same scale;
velocity runs from -200 to +200 km s$^{-1}$ (local standard of rest)
on each of the spectra, and the intensity scale is zero to 150 mJy per
beam for each.  The area of the grey scale map is considerably larger than the
ATCA primary beam (fwhm = 30'); the apparent concentration of the
polarized continuum flux to the central part of the map is a result
of the attenuation due to the primary beam.

Figure 3.  Spectra of Q' (see text) toward the middle
two squares in the second row on figure 2.  The dashed curves
indicate the noise level ($\pm 1 \sigma$ and $\pm 3 \sigma$) 
as measured in Stokes V, which 
includes the contribution to the system temperature from
the $\lambda$21-cm emission in the primary beam.


\begin{references}

\reference{Beck96}
Beck, R., Brandenburg, A., Moss, D., Shukurov, A., Sokoloff, D., 1996,
\araa 34, 155.

\reference{Brand93}
Brand, J., and Blitz, L., 1993, \aap 275, 67.

\reference{Burn66}
Burn, B.J., 1966, \mnras 133, 67.

\reference{Burton88}
Burton, W.B., 1988, in Galactic and Extragalactic Radio Astronomy, 2nd ed., eds.
 G.L. Verschuur and K. Kellerman, (New York :  Springer-Verlag) p. 295.

\reference{Dickey90}
Dickey, J.M. and Lockman, F.J., 1990, \araa 28, 215.

\reference{Duncan97}
Duncan, A.R., Haynes, R.F., Jones, K.L., and Stewart, R.T., 1997, \mnras in press.

\reference{Ehle93}
Ehle, M. and Beck, R., 1993, \aap 273, 45.
	
\reference{Ehle96}
Ehle, M.; Beck, R.; Haynes, R. F.; Vogler, A.; Pietsch, W.; Elmoutti, M.; Ryder,
 S., 1996, \aap 306, 73.

\reference{Han97}
Han, J.L., Manchester, R.N., Berkjuijsen, E.M., and Beck, R., 1997, \aap in press.

\reference{Junkes87}
Junkes, N., F\"{u}rst, E., and Reich, W., 1987, \aaps, 69, 451.

\reference{Klein96}
Klein, U., Hummel, E., Bomans, D.J., and Hopp, U., 1996, \aap 313, 396.

\reference{Liszt93}
Liszt, H.S., Braun, R., and Greisen, E., 1993, \aj 106, 2349.

\reference{Rand94}
Rand, R.J. and Lyne, A.G., 1994, \mnras 268, 497.

\reference{Salter88}
Salter, C.J. and Brown, R.L., 1988, in Galactic and Extragalactic Radio Astronomy, 
2nd ed., eds.  G.L. Verschuur and K. Kellerman, (New York :  Springer-Verlag) p. 1.

\reference{Spoelstra84}
Spoelstra, T.A.T., 1984, \aap 135, 238. 

\reference{Troland89}
Troland, T.H., Heiles, C., and Goss, W.M., 1989, \apj. 337, 342.

\reference{Whiteoak96}
Whiteoak, J.B.Z., and Green, A.J., 1996, \aaps 118, 329.

\reference{Wieringa93}
Wieringa, M.H., de Bruyn, A.G., Jansen, D., Brouw, W.N., and Katgert, P.,
1993, \aap, 268, 215.
\end{references}
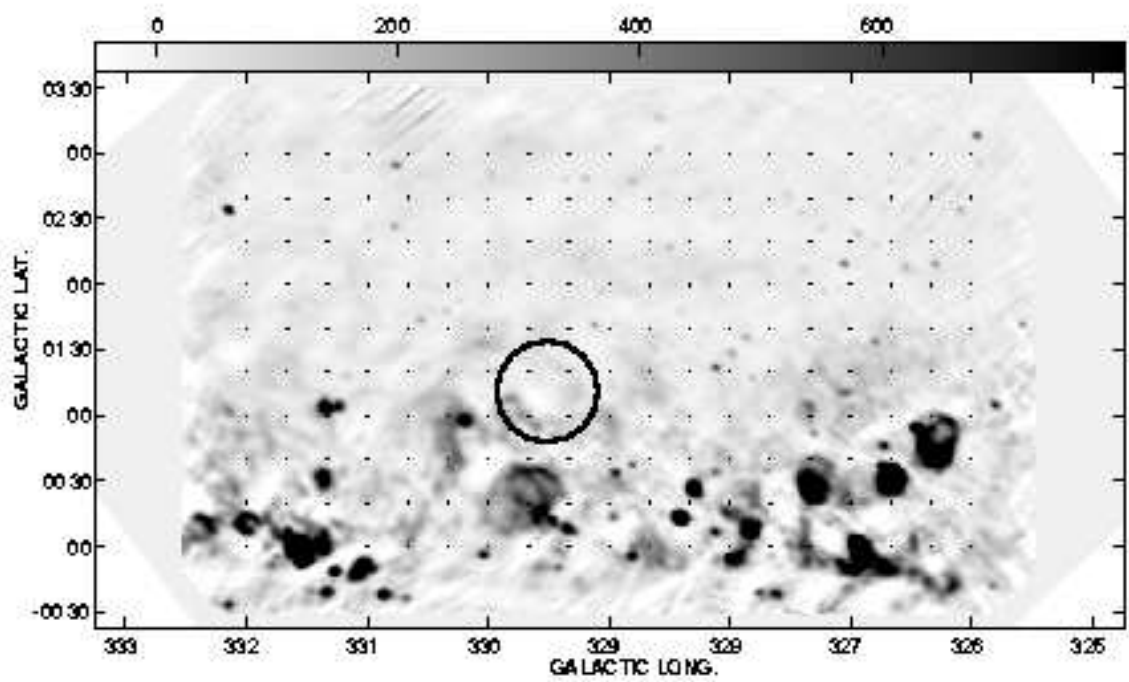
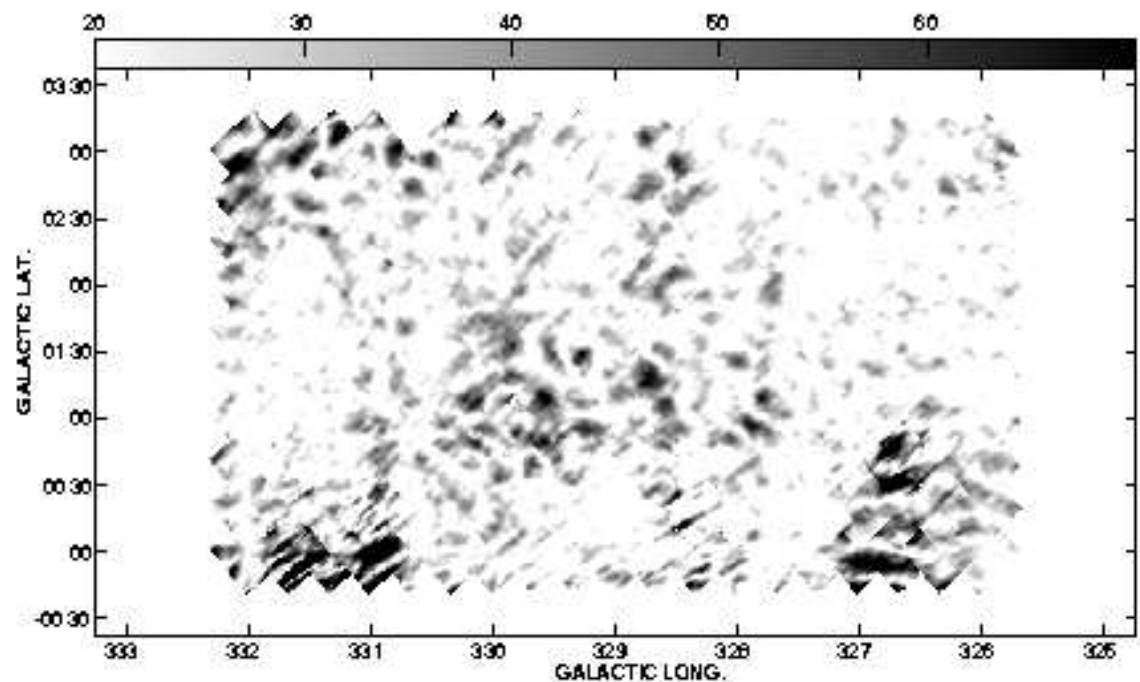
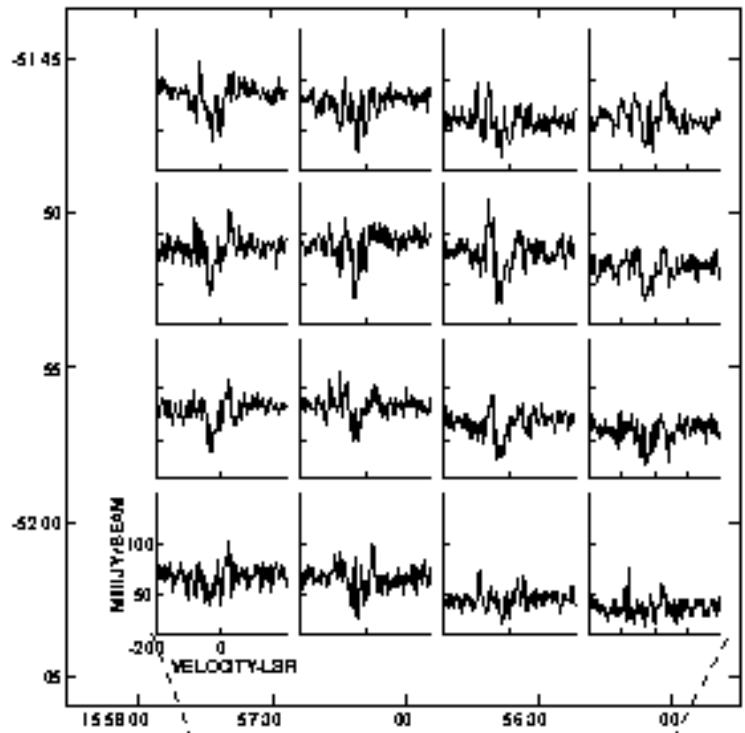
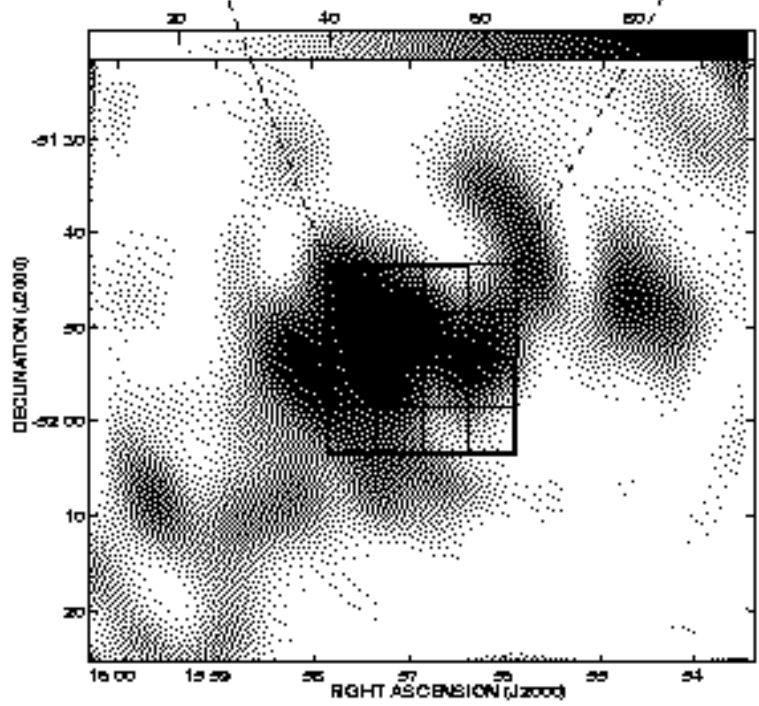
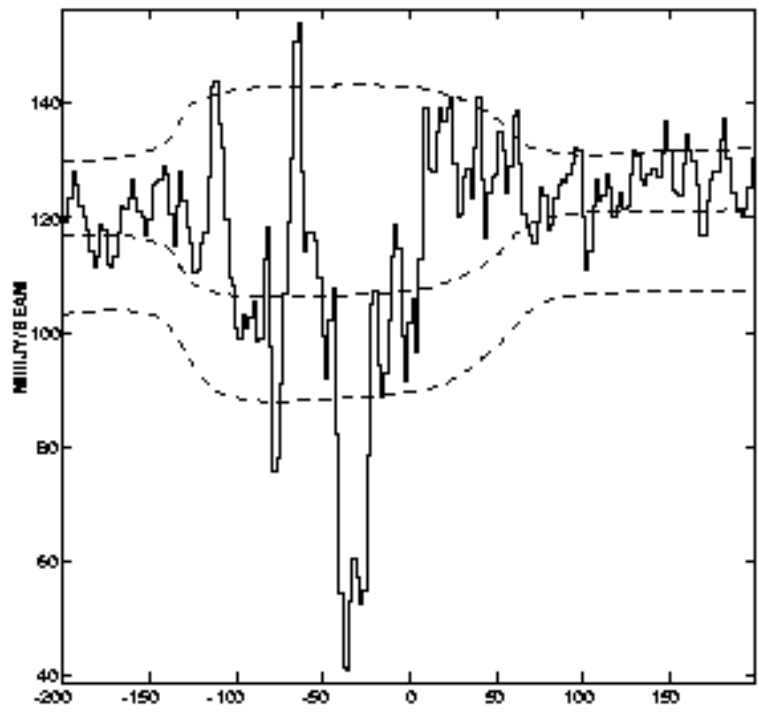
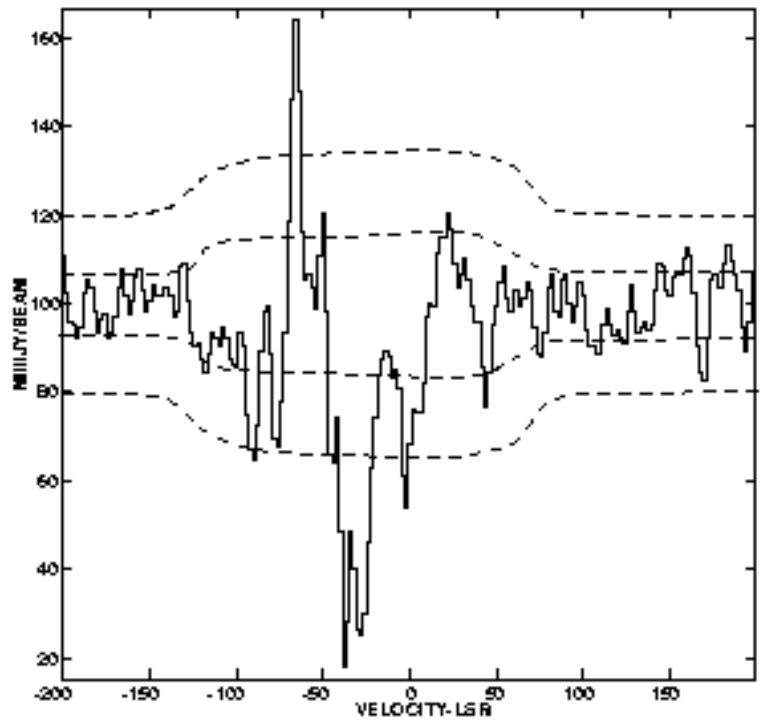
\end{document}